\documentclass{appolb}
\usepackage{amsmath,amssymb,amsfonts}
\begin{document}

\title{Nonlinear Field Space Theory and Quantum Gravity\thanks{Presented at {\it The 3rd Conference of the Polish Society on Relativity}, Krak\'{o}w, Poland, September 26-29, 2016.}
}

\author{Tomasz Trze\'{s}niewski\footnote{tbwbt@ift.uni.wroc.pl}
\address{Institute for Theoretical Physics, University of Wroc\l{}aw, pl.\ Borna 9, \\50-204 Wroc\l{}aw, Poland \\Institute of Physics, Jagiellonian University, {\L}ojasiewicza 11, \\30-348 Krak\'{o}w, Poland}}

\date{January 23, 2017}

\maketitle

\begin{abstract}
Phase spaces with nontrivial geometry appear in different approaches to quantum gravity and can also play a role in e.g.\! condensed matter physics. However, so far such phase spaces have only been considered for particles or strings. We propose an extension of the usual field theories to the framework of fields with nonlinear phase space of field values, which generally means nontrivial topology or geometry. In order to examine this idea we construct a prototype scalar field with the spherical phase space and then study its quantized version with the help of perturbative methods. As the result we obtain a variety of predictions that are known from the quantum gravity research, including algebra deformations, generalization of the uncertainty relation and shifting of the vacuum energy.
\end{abstract}

\section{Introduction}
M. Born \cite{Born:1938} was the first to suggest that, in the quantum theory of gravity, curved geometry of spacetime should be accompanied by momentum space that is similarly nontrivial. Indeed, curved momentum spaces or, in general, phase spaces with nontrivial geometry have been considered in different models of quantum gravity \cite{AmelinoCamelia:2011,Cianfrani:2014,Girelli:2010,Bojowald:2011} and this often leads to specific phenomenological predictions. The most rigorous approach is formulated in the language of quantum (Hopf) algebras. However, the discussed phase spaces are usually understood as belonging to some real or test particles. In \cite{Mielczarek:2016} we proposed to generalize this notion to the domain of field theory, so that the phase space of values of a given field is a nontrivial manifold, which becomes a linear (i.e.\! affine) space only in a certain limit. Similar constructions were already known in the context of string theory, where strings are described by non-linear sigma models, which (in the Tseytlin formulation) can be interpreted as having curved phase spaces \cite{Freidel:2014}. Another motivation comes from the principle of finiteness of physical quantities, which has been the idea behind the Born-Infeld theory \cite{Born:1934}, where field values are constrained by its dynamics. In our case such a constraint can be imposed a priori, by choosing a compact phase space.

\section{Toy model with the spherical phase space}
As a basic example \cite{Mielczarek:2016} we take a (massless) scalar field on $\mathbb{R}^{3,1}$. Its classical Hamiltonian 
in the Fourier representation has the form
\begin{align}\label{eq:02}
H = \tfrac{1}{2} \sum_{\bf k} \left( \pi_{\bf k}^2 + k^2 \phi_{\bf k}^2 \right), \quad k \equiv \sqrt{\bf k \cdot k}\,.
\end{align}
The phase space of each mode ${\bf k}$ is $\Gamma_{\bf k} := T^*(\mathbb{R}) = \mathbb{R}^2 \ni (\phi_{\bf k},\pi_{\bf k})$, with the Poisson bracket $\{\phi_{\bf k}, \pi_{\bf k}\} = 1$, and the total field phase space $\Gamma = \prod_{\bf k} \Gamma_{\bf k}$. 

Let us now assume that $\forall_{\bf k}: \Gamma_{\bf k} = S^2$ (a sphere). On $S^2$, covered by angular coordinates $\varphi$, $\theta$, the natural symplectic form is given by the area form $\omega = J \sin\theta\, d\varphi \wedge d\theta$, $\int_{S^2} \omega = 4\pi J$, with the non-linearity scale $J$. We accordingly parametrize field variables $\phi_{\bf k}$, $\pi_{\bf k}$ in terms of $\varphi$, $\theta$ as
\begin{align}\label{eq:03}
R^{-1} \phi_{\bf k} = \varphi - \pi \in [-\pi, \pi)\,, \qquad R J^{-1} \pi_{\bf k} = \tfrac{\pi}{2} - \theta \in \left[-\tfrac{\pi}{2}, \tfrac{\pi}{2}\right],
\end{align}
where $R$ denotes a dimensionful constant and (for simplicity) we choose $J$ to be ${\bf k}$-independent. The symplectic form becomes $\omega = \cos(\frac{R}{J} \pi_{\bf k})\, d\pi_{\bf k} \wedge d\phi_{\bf k}$ and the corresponding Poisson bracket
\begin{align}\label{eq:05}
\left\{\phi_{\bf k}, \pi_{\bf k}\right\} = \sec\left(R J^{-1} \pi_{\bf k}\right).
\end{align}
The same construction can be made \cite{Mielczarek:2017} for points of $\mathbb{R}^{3,1}$ instead of modes. 

Since the variables $\phi_{\bf k}$, $\pi_{\bf k}$ in (\ref{eq:03}) are not everywhere well defined, it is often convenient to switch to the spin-like coordinates
\begin{align}\label{eq:07}
J_{(x)} := J \sin\theta \cos\varphi\,, \qquad 
J_{(y)} := J \sin\theta \sin\varphi\,, \qquad 
J_{(z)} := J \cos\theta\,,
\end{align}
satisfying the relation $J_{(x)}^2 + J_{(y)}^2 + J_{(z)}^2 = J^2$, where $\phi$, $\theta$ are expressed through the formulae (\ref{eq:03}). Calculating the brackets (\ref{eq:05}) for $J_i$'s one verifies that they span the usual $\mathfrak{su}(2)$ Lie algebra $\{J_i, J_j\} = \epsilon_{ijk} J^k$. 

Similarly, in order to find a Hamiltonian which is globally well defined, has the usual minimum $(\phi_{\bf k},\pi_{\bf k}) = (0,0)$ and correct linearized limit we may apply the formal analogy with a spin ${\bf J}$ in a constant magnetic field ${\bf B}$ and postulate the Hamiltonian of the form (analogous to $H \sim {\bf B} \cdot {\bf J}$)
\begin{align}\label{eq:08}
H = \sum_{\bf k} H_{\bf k}\,, \qquad H_{\bf k} := k J_{(x)} = -J k \cos\frac{\pi_{\bf k}}{\sqrt{J k}}\, \cos\frac{\phi_{\bf k}}{\sqrt{J/k}}\,, 
\end{align}
where we also fixed $R = \sqrt{J/k}$. The ordinary Hamiltonian (\ref{eq:02}), up to a energy spectrum shift by $-Jk$, is recovered in the limit $J \rightarrow \infty$. Calculating the brackets $\dot{f} = \left\{f, H_{\bf k}\right\}$, $f = \phi_{\bf k},\pi_{\bf k}$ we obtain the Hamilton equations
\begin{align}\label{eq:09}
\dot{\phi}_{\bf k} = {\sqrt{J k}}\, \tan\frac{\pi_{\bf k}}{\sqrt{J k}}\, \cos\frac{\phi_{\bf k}}{\sqrt{J/k}}\,, \qquad 
\dot{\pi}_{\bf k} = -\sqrt{J k}\, k \sin\frac{\phi_{\bf k}}{\sqrt{J/k}}\,,
\end{align}
which describe phase space trajectories\footnote{The form of solutions (\ref{eq:10}) was found during the conference by I. Bia\l ynicki-Birula, whom we thank for his interest.} with parameters $C,t_0 \in \mathbb{R}$:
\begin{align}\label{eq:10}
\phi_{\bf k}(t) = \sqrt{J/k} \arcsin\left(C \cos(k (t - t_0))/\sqrt{J/k - C^2 \sin^2(k (t - t_0))}\right), \nonumber\\
\pi_{\bf k}(t) = -\sqrt{J k} \arcsin\left(C \sqrt{k/J} \sin(k (t - t_0))\right).
\end{align}
Each solution (\ref{eq:10}) outlines a circle with the center at $\phi_{\bf k},\pi_{\bf k} = 0$ but for the great circle (i.e.\! when $C = \pm\sqrt{J/k}$) they become singular and therefore cover only half of the sphere. In the limit $J \rightarrow \infty$ we recover the classical expressions $\phi_{\bf k}(t) = C \cos(k (t - t_0))$, $\pi_{\bf k}(t) = C k \sin(k (t - t_0))$. 


\section{Selected results from the quantized model}
Inspired by the polymer quantization approach \cite{Bojowald:2011}, we assume \cite{Mielczarek:2016} that the quantum version of the bracket (\ref{eq:05}) is given by the $\mathfrak{su}(2)$ commutator $[\hat{J}_i, \hat{J}_j] = i \hslash\, \epsilon_{ijk} \hat{J}^k$. Consequently, our phase space can not be globally decomposed into field values and momenta and to represent it one has to use quasiprobability distributions, such as the Wigner function, instead of usual wave functions. Nevertheless, for quantum states supported on field values $\phi_{\bf k} \ll \frac{\pi}{2} \sqrt{J/k}$, $\pi_{\bf k} \ll \frac{\pi}{2} \sqrt{J k}$ we can expand $\hat{J}_i$'s in terms of $\hat{\phi}_{\bf k},\hat{\pi}_{\bf k}$ and derive the deformed commutation relation
\begin{align}\label{eq:13}
[\hat{\phi}_{\bf k},\hat{\pi}_{\bf k}] \approx i \hslash \left( \hat{\mathbb{I}} - \frac{k}{2J} \hat{\phi}_{\bf k}^2 - \frac{1}{2Jk} \hat{\pi}_{\bf k}^2 \right).
\end{align}
It naturally corresponds to the generalized uncertainty principle
\begin{align}\label{eq:14}
\Delta \hat{\phi}_{\bf k} \Delta \hat{\pi}_{\bf k} \geq \frac{\hslash}{2} \left( 1 - \frac{k}{2J} (\Delta \hat{\phi}_{\bf k})^2 - \frac{1}{2Jk} (\Delta \hat{\pi}_{\bf k})^2 \right)
\end{align}
(if the mean values $\langle \hat{\phi}_{\bf k} \rangle, \langle \hat{\pi}_{\bf k} \rangle = 0$), which may be compared with e.g.\! \cite{Bojowald:2011}. 

Furthermore, keeping the expansion to the order $J^{-1}$, we can express $\hat{\phi}_{\bf k}$ and $\hat{\pi}_{\bf k}$ in terms of the creation and annihilation operators $\hat{a}_{\bf k}^\dagger$, $\hat{a}_{\bf k}$ as
\begin{align}\label{eq:15}
\hat{\phi}_{\bf k} = \sqrt{\frac{\hslash J}{(\hslash + 2J) k}} \left( \hat{a}_{\bf k}^\dagger + \hat{a}_{\bf k} \right), \qquad \hat{\pi}_{\bf k} = i \sqrt{\frac{\hslash J k}{\hslash + 2J}} \left( \hat{a}_{\bf k}^\dagger - \hat{a}_{\bf k} \right).
\end{align}
$\hat{a}_{\bf k}^\dagger$, $\hat{a}_{\bf k}$ then generate a $Q$-deformed oscillator algebra $\hat{a}_{\bf k} \hat{a}^\dagger_{\bf k} - Q\, \hat{a}^\dagger_{\bf k} \hat{a}_{\bf k} = \hat{\mathbb{I}}$, where the deformation parameter $Q \equiv (1 - \frac{\hslash}{2J})/(1 + \frac{\hslash}{2J}) = 1 - \frac{\hslash}{J} + \mathcal{O}(J^{-2})$. 
Subsequently, the quantized Hamiltonian (\ref{eq:08}), i.e.\! $\hat{H}_{\bf k} := k \hat{J}_{(x)}$ with the symmetric ordering of $\hat{\phi}_{\bf k}$ and $\hat{\pi}_{\bf k}$, can be perturbatively expanded in $J^{-1}$. As the result we find the energy eigenvalues (with $n \in \mathbb{N}_0$)
\begin{align}\label{eq:18}
E_n = -Jk + \hslash k \left( n + \tfrac{1}{2} \right) - \tfrac{1}{4} J^{-1} \hslash^2 k \left( 3n^2 + 3n + 1 \right) + \mathcal{O}(J^{-2})
\end{align}
and the corresponding eigenstates
\begin{align}\label{eq:19}
|n \rangle = |n^{(0)} \rangle + c_{n+4} |(n+4)^{(0)} \rangle  
+ c_{n-4} |(n-4)^{(0)} \rangle\big\vert_{n\geq 4} + \mathcal{O}(J^{-2})\,,
\end{align}
where the index $(0)$ denotes the zeroth order of the expansion, while the coefficients $c_{n+4} \equiv -\frac{\hslash}{96 J} \sqrt{(n+4)!/n!}$, $c_{n-4} \equiv \frac{\hslash}{96 J} \sqrt{n!/(n-4)!}$. In particular, the standard vacuum energy $E_0 = \frac{1}{2} \hslash k$ is shifted by $-J k - \frac{1}{4J} \hslash^2 k$.

The properties of the toy model discussed here and in \cite{Mielczarek:2016} show the potential usefulness of our framework in the context of quantum gravity. It can also be tested in cosmology (J.~Mielczarek and T.T., in preparation).\\

I acknowledge the support by the Polish Ministry of Science and Higher Education, project 0302/IP3/2015/73 and by the National Science Centre Poland, project 2014/13/B/ST2/04043.

\end{document}